\documentclass[aps,prc,preprint,superscriptaddress,showpacs]{revtex4}
\usepackage[dvips]{graphicx}
\usepackage{amsmath,amssymb}
\usepackage{ulem}

\newcommand{\sik}{\hspace{0.3em}\raisebox{0.4ex}{$<$}
            \hspace{-0.73em}\raisebox{-.7ex}{$\sim$}\hspace{0.3em}}
\newcommand{\beq}{\begin{equation}}
\newcommand{\eeq}{\end{equation}}
\newcommand{\bea}{\begin{eqnarray}}
\newcommand{\eea}{\end{eqnarray}}

\DeclareSymbolFont{boldletters}{OML}{cmm} {b}{it}
\DeclareSymbolFontAlphabet{\mathbit}{boldletters}
\DeclareMathSymbol{\alpha}{\mathalpha}{letters}{"0B}
\DeclareMathSymbol{\beta}{\mathalpha}{letters}{"0C}
\DeclareMathSymbol{\gamma}{\mathalpha}{letters}{"0D}
\DeclareMathSymbol{\delta}{\mathalpha}{letters}{"0E}
\DeclareMathSymbol{\epsilon}{\mathalpha}{letters}{"0F}
\DeclareMathSymbol{\zeta}{\mathalpha}{letters}{"10}
\DeclareMathSymbol{\eta}{\mathalpha}{letters}{"11}
\DeclareMathSymbol{\theta}{\mathalpha}{letters}{"12}
\DeclareMathSymbol{\iota}{\mathalpha}{letters}{"13}
\DeclareMathSymbol{\kappa}{\mathalpha}{letters}{"14}
\DeclareMathSymbol{\lambda}{\mathalpha}{letters}{"15}
\DeclareMathSymbol{\mu}{\mathalpha}{letters}{"16}
\DeclareMathSymbol{\nu}{\mathalpha}{letters}{"17}
\DeclareMathSymbol{\xi}{\mathalpha}{letters}{"18}
\DeclareMathSymbol{\pi}{\mathalpha}{letters}{"19}
\DeclareMathSymbol{\rho}{\mathalpha}{letters}{"1A}
\DeclareMathSymbol{\sigma}{\mathalpha}{letters}{"1B}
\DeclareMathSymbol{\tau}{\mathalpha}{letters}{"1C}
\DeclareMathSymbol{\upsilon}{\mathalpha}{letters}{"1D}
\DeclareMathSymbol{\phi}{\mathalpha}{letters}{"1E}
\DeclareMathSymbol{\chi}{\mathalpha}{letters}{"1F}
\DeclareMathSymbol{\psi}{\mathalpha}{letters}{"20}
\DeclareMathSymbol{\omega}{\mathalpha}{letters}{"21}
\DeclareMathSymbol{\varepsilon}{\mathalpha}{letters}{"22}
\DeclareMathSymbol{\vartheta}{\mathalpha}{letters}{"23}
\DeclareMathSymbol{\varpi}{\mathalpha}{letters}{"24}
\DeclareMathSymbol{\varrho}{\mathalpha}{letters}{"25}
\DeclareMathSymbol{\varsigma}{\mathalpha}{letters}{"26}
\DeclareMathSymbol{\varphi}{\mathalpha}{letters}{"27}
\DeclareMathSymbol{\Gamma}{\mathalpha}{letters}{"00}
\DeclareMathSymbol{\Delta}{\mathalpha}{letters}{"01}
\DeclareMathSymbol{\Theta}{\mathalpha}{letters}{"02}
\DeclareMathSymbol{\Lambda}{\mathalpha}{letters}{"03}
\DeclareMathSymbol{\Xi}{\mathalpha}{letters}{"04}
\DeclareMathSymbol{\Pi}{\mathalpha}{letters}{"05}
\DeclareMathSymbol{\Sigma}{\mathalpha}{letters}{"06}
\DeclareMathSymbol{\Upsilon}{\mathalpha}{letters}{"07}
\DeclareMathSymbol{\Phi}{\mathalpha}{letters}{"08}
\DeclareMathSymbol{\Psi}{\mathalpha}{letters}{"09}
\DeclareMathSymbol{\Omega}{\mathalpha}{letters}{"0A}

\begin{document}
\preprint{SAGA-HE-237-07}
\title{
Critical endpoint in the Polyakov-loop extended NJL model}

\author{Kouji Kashiwa}
\email[]{kashiwa2scp@mbox.nc.kyushu-u.ac.jp}
\affiliation{Department of Physics, Graduate School of Sciences, Kyushu University,
             Fukuoka 812-8581, Japan}

\author{Hiroaki Kouno}
\email[]{kounoh@cc.saga-u.ac.jp}
\affiliation{Department of Physics, Saga University,
             Saga 840-8502, Japan}

\author{Masayuki Matsuzaki}
\email[]{matsuza@fukuoka-edu.ac.jp}
\affiliation{Department of Physics, Fukuoka University of Education, 
             Munakata, Fukuoka 811-4192, Japan}

\author{Masanobu Yahiro}
\email[]{yahiro2scp@mbox.nc.kyushu-u.ac.jp}
\affiliation{Department of Physics, Graduate School of Sciences, Kyushu University,
             Fukuoka 812-8581, Japan}

\date{\today}

\begin{abstract}
The critical endpoint (CEP) and the phase structure
are studied in the Polyakov-loop extended Nambu--Jona-Lasinio model in which
the scalar type eight-quark ($\sigma^4$) interaction 
and the vector type four-quark interaction are newly added. 
The $\sigma^4$ interaction largely shifts the CEP 
toward higher temperature and lower chemical potential, 
while the vector type interaction does oppositely. 
At zero chemical potential, 
the $\sigma^4$ interaction moves the pseudo-critical
temperature of 
the chiral phase transition to the vicinity of that of the deconfinement phase transition. 
\end{abstract}

\pacs{11.30.Rd, 12.40.-y}
\maketitle


The position of the critical endpoint (CEP) in the 
phase diagram is 
one of the most interesting subjects in hot and dense Quantum Chromodynamics (QCD). 
With the aid of the progress in computer power, 
lattice QCD simulations have become feasible for thermal systems 
at zero chemical potential ($\mu$)~\cite{Kog}. 
For finite chemical potential, however, 
lattice QCD has the well-known sign problem,
so that only a few
works were made to determine the position of the CEP~\cite{ZF,Ejiri}. 

As an approach complementary to first-principle 
lattice simulations, one can consider several effective models. 
One of them is the Nambu--Jona-Lasinio (NJL) model~\cite{NJ1}.
In the original NJL model that includes 
scalar and pseudo-scalar type four-quark interactions, 
it was found that there exists a CEP in the phase diagram~\cite{AY,Sca}. 
However, the CEP is located at a lower temperature ($T$) and a higher 
$\mu$ compared with 
the one predicted by a lattice QCD simulation~\cite{ZF} 
and by the QCD-like theory~\cite{KMT,HTF}. 
Moreover, recent empirical analysis~\cite{Lacey}  
of $\eta/s$, the ratio of 
shear viscosity to entropy density,  
suggest there is a CEP
at $T_{\rm e} \sim 165$ MeV and $\mu_{\rm e} \sim 50 - 60$ MeV
that is much higher $T$ and lower $\mu$ than the NJL model predictions.

Kashiwa et al.~\cite{Kashiwa1} showed that in the NJL model 
the scalar type eight-quark ($\sigma^4$) interaction newly added shifts the CEP 
toward higher $T$ and lower $\mu$. 
However, the location is still far from predictions of 
lattice QCD, the QCD-like theory and the empirical analyses. 

It is also known in the chiral hadron model~\cite{ZZSS,ZZS} 
that a CEP appears at $T$ much higher than the prediction of the NJL model. 
This implies that the deconfinement phase transition plays an important role in determining the position of the CEP. 
Although the NJL model is a useful method 
for understanding the chiral symmetry breaking, 
this model does not possess a confinement mechanism.
As a reliable model that can treat both the chiral and the deconfinement phase transitions,  
we can consider the Polyakov-loop extended NJL (PNJL) model
~\cite{Meisinger,Fukushima,Ghos,Megias,Ratti1,Ratti2,Rossner,
 Hansen,Sasaki1,Schaefer}.
In the PNJL model the confinement/deconfinement phase
transition is described by the Polyakov loop.
Effects of the Polyakov loop 
make the CEP move to higher $T$ and lower $\mu$ than the NJL model predicts~\cite{Rossner}. 
The position of the CEP is still far from the predictions of lattice QCD 
and the empirical analyses.

Meanwhile, it was recently reported that the vector type interaction~\cite{Buballa,KKKN,Kashiwa1,Kashiwa2} 
is necessary to realize the heavy neutron star~\cite{Blaschke}. 
This may indicate that the vector type interaction is necessary in the finite $\mu$ region of the  phase diagram. 
  
In this letter, we study effects of the $\sigma^4$ and the vector type interactions 
on the position of the CEP and the interplay between
the chiral and the deconfinement phase transitions, by using the PNJL model with these interactions. 

The model we consider here is 
the following two-flavor PNJL model 
with the vector type interaction~\cite{Buballa,KKKN,Kashiwa1,Kashiwa2} 
and the $\sigma^4$  
interaction~\cite{Osipov1,Osipov2,Osipov3,Osipov4,Kashiwa1} 
\begin{eqnarray}
 {\cal L} &=& {\bar q}(i \gamma_\mu D^\mu -m_0)q + {\cal L}_\mathrm{int} 
             - {\cal U}(\Phi [A],\bar{\Phi} [A],T) ,
\label{eq:E1}
\end{eqnarray}
where $q$ denotes the two-flavor quark field, 
$m_0$ the current quark mass and $D^\mu=\partial^\mu-iA^\mu$ the covariant derivative. 
The field $A^\mu$ is defined as 
$ A^\mu=\delta_{\mu0}gA^\mu_a{\lambda^a\over{2}}$
with the gauge field $A^\mu_a$, the Gell-Mann matrix $\lambda_a$ and the gauge coupling $g$.
The interaction of the NJL sector, ${\cal L}_\mathrm{int}$, is
\begin{eqnarray}
{\cal L}_\mathrm{int} &=&   G_{\rm s}[({\bar q}q)^2 
                   + ({\bar q}i\gamma_5 {\vec \tau}q)^2]
                   + G_{\rm s8}[({\bar q}q)^2 
                   + ({\bar q}i\gamma_5 {\vec \tau}q)^2]^2
                   - G_{\rm v}({\bar q}\gamma^\mu q)^2, \label{eq:E4}
\label{eq:E5-2}
\end{eqnarray}
where 
${\vec \tau}$ stands for the isospin matrix, and  
$G_{\rm s}$, $G_{\rm v}$ and $G_{\rm s8}$  
denote coupling constants of the scalar type 
four-quark, the vector type four-quark and  
the $\sigma^4$  interactions, respectively. 
The Polyakov potential ${\cal U}$, defined in Eq. (\ref{eq:E13}), 
is a function of the Polyakov loop $\Phi$ and its conjugate $\bar \Phi$,
\begin{eqnarray}
\Phi        &=& {1\over{N_{\rm c}}}({\rm Tr} L),~~~~
{\bar \Phi}  = {1\over{N_{\rm c}}}({\rm Tr}L^\dag),
\end{eqnarray}
with
\begin{eqnarray}
L({\bf x})  &=& {\cal P} \exp\Bigl[
                {i\int^\beta_0 d \tau A_4({\bf x},\tau)}\Bigr],
\end{eqnarray}
where ${\cal P}$ is the path ordering and $A_4 = iA_0 $.
In the chiral limit ($m_0=0$), 
the Lagrangian density has exact 
$SU(2)_{\rm L} \times SU(2)_{\rm R}
\times U(1)_{\rm v} \times SU(3)_{\rm c}$  symmetry.

The temporal component of the gauge field is diagonal 
in the flavor space, because the color and the flavor spaces 
are completely separated out in the present case. 
In the Polyakov gauge,
the Polyakov loop matrix $L$ can be written in a diagonal form 
in the color space~\cite{Fukushima}: 
\begin{eqnarray}
\frac{1}{N_{\rm c}} ({\rm Tr} L) 
      &=& \frac{1}{N_{\rm c}}({\rm Tr}~e^{i \beta (\phi_3 \lambda_3 + \phi_8 \lambda_8)}),\\
      &=& \frac{1}{N_{\rm c}}\Bigl({\rm Tr}~{\rm diag} (e^{i \beta \phi_{\rm a}},e^{i \beta \phi_{\rm b}},
e^{i \beta \phi_{\rm c}} )\Bigr),
\label{eq:E6}
\end{eqnarray}
where $\phi_{\rm a}=\phi_3+\phi_8/\sqrt{3}$, $\phi_{\rm b}=-\phi_3+\phi_8/\sqrt{3}$
and $\phi_{\rm c}=-(\phi_{\rm a}+\phi_{\rm b})=-2\phi_8/\sqrt{3}$. 
The Polyakov loop is an exact order parameter of the spontaneous 
${\mathbb Z}(N_\mathrm{c})$ symmetry breaking in the pure gauge theory.
Although ${\mathbb Z}(N_\mathrm{c})$ is not an exact symmetry in the system with dynamical quarks, 
it still seems to be a good indicator of 
the deconfinement phase transition. 
Therefore, we use $\Phi$ to define the deconfinement phase transition.

Under the mean field approximation (MFA), the Lagrangian density becomes
\begin{eqnarray}
{\cal L}_{\rm MFA} &=& {\bar q}( i \gamma_\mu D^\mu - (m_0+\Sigma_{\rm s}) 
                      + \Sigma_{\rm v} \gamma^0)q 
                      - U(\sigma,\rho_{\rm v}) - {\cal U}(\Phi,{\bar \Phi},T), 
\label{eq:E7} 
\end{eqnarray}
where
\begin{eqnarray}
\sigma &=& \langle \bar{q}q \rangle, ~~~~~
\rho_{\rm v}(T,\mu,\sigma,\Phi,{\bar \Phi})=\langle \bar{q}\gamma_0q \rangle, 
\label{eq:E8} \\
\Sigma_{\rm s} &=& -(2 G_{\rm s} \sigma + 4 G_{\rm s8}\sigma^3), 
~~~~~\Sigma_{\rm v}=-2 G_{\rm v} \rho_{\rm v},\label{eq:E9} \\
U        &=&  G_{\rm s} \sigma^2 + 3G_{\rm s8} \sigma^4 
              - G_{\rm v}\rho_{\rm v}^2. \label{eq:E10}
\end{eqnarray}
In the $1/N_\mathrm{c}$ expansion,
the eight-quark interaction after the MFA, ${\bar q} \Gamma q \langle {\bar q} \Gamma^\prime q \rangle^3$ 
is of order $N^0_\mathrm{c}$ and accordingly the same order as the
four-quark interaction after the MFA, ${\bar q} \Gamma q \langle {\bar
q} \Gamma^\prime q \rangle$, where $\Gamma$ and $\Gamma^\prime$ 
are vertex matrices. 
Therefore, we can not ignore higher multi-quark interactions in general. 
As a starting point, we take into account the scalar type 
eight-quark ($\sigma^4$) interaction that is known to affect the
position of the CEP \cite{Kashiwa1}. 
Using the usual techniques~\cite{Kapusta,Bellac}, 
one can obtain the thermodynamical potential
\begin{eqnarray}
\Omega&=&-T\ln Z\\
      &=& -2 N_{\rm f} V \int \frac{d^3{\rm p}}{(2\pi)^3}
         \Bigl[ 3 E ({\rm p}) + \frac{1}{\beta}
         \bigl\{
        {\rm Tr_c}\ln(1+Le^{-\beta E^-({\rm p})}
        )  \nonumber\\
      &&  + {\rm Tr_c}\ln(1+L^\dag e^{-\beta E^+({\rm p})})
        \bigr\}
        \Bigr] +\Bigl\{U(\sigma,\omega)+{\cal U}({\bar \Phi},\Phi,T)\Bigr\}V, 
\label{eq:E11}
\end{eqnarray}
where $E({\rm p})=\sqrt{{\bf p}^2+M^2}$, 
$E^\pm=E({\rm p})\pm {\tilde \mu}$ ,$M=m_0 + \Sigma_{\rm s}$ and
${\tilde \mu} = \mu + \Sigma_{\rm v}$. 
After some algebra, the thermodynamical potential $\Omega$ becomes~\cite{Ratti1}
\begin{eqnarray}
\Omega&=& -2 N_{\rm f}V \int \frac{d^3{\rm p}}{(2\pi)^3}
         \Bigl[ 3 E ({\rm p}) \nonumber\\
      && + \frac{1}{\beta}
           \ln~ [1 + 3(\Phi+{\bar \Phi e^{-\beta E^-({\bf p})}}) 
           e^{-\beta E^-({\bf p})}+ e^{-3\beta E^-({\bf p})}]
         \nonumber\\
      && + \frac{1}{\beta} 
           \ln~ [1 + 3({\bar \Phi}+{\Phi e^{-\beta E^+({\bf p})}}) 
              e^{-\beta E^+({\bf p})}+ e^{-3\beta E^+({\bf p})}]
	      \Bigl]\nonumber\\
      && +(U+{\cal U})V. 
\label{eq:E12} 
\end{eqnarray}

In the $T=0$ limit, the PNJL model 
is reduced to the NJL model, since 
the Polyakov loop is included only in $\cal U$ that has no $\sigma$ dependence, as shown in 
\begin{eqnarray}
\Omega &=& -6N_{\rm f}V \int \frac{d^3{\rm p}}{(2\pi)^3}
           \Bigl[E ({\bf p}) -\theta (-E^- ({\bf p}))E^- ({\bf p}) \Bigl]
\nonumber\\
           &&+\Bigl[U(\sigma ,\rho_{\rm v}(T\to 0,\mu,\sigma ))+{\cal U}(T\to 0,\Phi,\bar{\Phi})\Bigl]V. 
\label{eq:E12-2} 
\end{eqnarray}

We use ${\cal U}$ of Ref.~\cite{Ratti1} that is fitted to a lattice QCD simulation 
in the pure gauge theory at finite $T$~\cite{Boyd,Kaczmarek}: 
\begin{eqnarray}
{{\cal U}\over{T^4}} &=&  -\frac{b_2(T)}{2} {\bar \Phi}\Phi
              -\frac{b_3}{6}({\bar \Phi^3}+ \Phi^3)
              +\frac{b_4}{4}({\bar \Phi}\Phi)^2, \label{eq:E13} \\
b_2(T)   &=& a_0 + a_1\Bigl(\frac{T_0}{T}\Bigr)
                 + a_2\Bigl(\frac{T_0}{T}\Bigr)^2
                 + a_3\Bigl(\frac{T_0}{T}\Bigr)^3,  \label{eq:E14}
\end{eqnarray}
where parameters are summarized in Table I.  

The Polyakov potential yields a deconfinement phase transition at 
$T=T_0$ in the pure gauge theory.
Hence, $T_0$ is taken to be $270$ MeV predicted by the pure gauge lattice QCD calculation.

\begin{table}[h]
\begin{center}
\begin{tabular}{llllll}
\hline
~~~~~$a_0$~~~~~&~~~~~$a_1$~~~~~&~~~~~$a_2$~~~~~&~~~~~$a_3$~~~~~&~~~~~$b_3$~~~~~&~~~~~$b_4$~~~~~
\\
\hline
~~~~6.75 &~~~~-1.95 &~~~~2.625 &~~~~-7.44 &~~~~0.75 &~~~~7.5 
\\
\hline
\end{tabular}
\caption{
Summary of the parameter set in the Polyakov sector
used in Ref.~\cite{Ratti1}. 
All parameters are dimensionless. 
}
\end{center}
\end{table}

Since the NJL model is nonrenormalizable, it is needed to 
introduce a cutoff in the momentum integration. 
In this study, we use the three-dimensional momentum cutoff 
\begin{equation}
\int \frac{d^3{\bf p}}{(2 \pi)^3}\to 
{1\over{2\pi^2}} \int_0^\Lambda p^2dp .
\label{eq:E15}
\end{equation}
Hence, the present model has five parameters 
$m_0$, $\Lambda$, $G_{\rm s}$, $G_{\rm s8}$, and $G_{\rm v}$ in the NJL sector. 
We simply assume $m_0=$ 5.5 MeV. 
In the case without the $\sigma^4$  interaction, 
we use $\Lambda =0.6315$ GeV and $G_{\rm s}=5.498$ GeV$^{-2}$, 
which reproduce the empirical values of the pion decay constant and the pion mass, 
$f_\pi=$ 93.3 MeV and  $M_\pi=$ 138 MeV. 
In the case with the $\sigma^4$ interaction, 
we take $\Lambda=0.6315$ GeV, 
$G_{\rm s}=$ 5.002 (5.276) 
GeV$^{-2}$ and $G_{\rm s8}\sigma_0^2=$ 0.2476 GeV$^{-2}$ (0.1109 GeV$^{-2}$), 
which reproduce the pion decay constant $f_\pi=$ 93.3 MeV, the pion mass $M_\pi=$ 138 MeV and 
the sigma meson mass $M_\sigma=$ 600 MeV (650 MeV)~\cite{Kashiwa1}. 
It should be noted that larger $G_{\rm s8}$ yields smaller $M_\sigma$.

For the vector coupling constant $G_{\rm v}$, 
we take three values, 0, 0.25$G_{\rm s}$ and 0.5$G_{\rm s}$. 
The case $G_{\rm v}=0.5G_{\rm s}$ is obtained by one gluon exchange calculation  in perturbative QCD~\cite{Hatsuda2}. 
The case $G_{\rm v}=$0.25$G_{\rm s}$ is obtained 
by the instanton-anti-instanton molecule model~\cite{KKKN}. 

Table II summarizes the parameter sets we take.  
\begin{table}[h]
\begin{center}
\begin{tabular}{llll}
\hline
~~~~model 
& $G_{\rm s}$ & $G_{{\rm s}8} \sigma_0^2$ 
\\
\hline
~~~~Original
& 5.498
& 0~~~
\\
~~~~Original + $\sigma^4$ 
& 5.002
& 0.2476
\\
~~~~Original + $\sigma^4$ ($M_\sigma = 650$ MeV) ~~~~
& 5.276 ~~~~
& 0.1109 ~~~~
\\  
\hline
\end{tabular}
\caption{Summary of the parameter sets of the NJL part. 
The coupling constants are shown in ${\rm  GeV}^{-2}$.
For all cases, we take $\Lambda =0.6315$~GeV and 
$\sigma_0\equiv\sigma (T=0,\mu =0)=-0.03023$~GeV$^3$.}
\end{center}
\end{table}

Stationary conditions for $\sigma$, $\rho_{\rm v}$, $\Phi$ and $\bar{\Phi}$ 
become 
\begin{eqnarray}
\frac{1}{V}
\frac{\partial \Omega}{\partial \sigma}
&=& G_{\rm s}^* \sigma + 6N_{\rm f} G_{\rm s}^*\times \nonumber\\
&&\int\frac{d^3 p}{(2\pi)^3}\Bigl\{ \frac{M}{E({\bf p})}
\Bigl\{1-\frac{h^{(-)}(\sigma,\Phi,{\bar \Phi};T,\mu)}
              {g^{(-)}(\sigma,\Phi,{\bar \Phi};T,\mu)}
        -\frac{h^{(+)}(\sigma,\Phi,{\bar \Phi};T,\mu)}
              {g^{(+)}(\sigma,\Phi,{\bar \Phi};T,\mu)}\Bigr\}=0,
\label{eq:E16}
\\
\frac{1}{V}
\frac{\partial \Omega}{\partial \rho_{\rm v}}
&=& -2G_{\rm v} \rho_{\rm v} + 12N_{\rm f} G_{\rm v} \times \nonumber\\
&&\int\frac{d^3 p}{(2\pi)^3}
\Bigl\{\frac{h^{(-)}(\sigma,\Phi,{\bar \Phi};T,\mu)}
            {g^{(-)}(\sigma,\Phi,{\bar \Phi};T,\mu)}
      -\frac{h^{(+)}(\sigma,\Phi,{\bar \Phi};T,\mu)}
            {g^{(+)}(\sigma,\Phi,{\bar \Phi};T,\mu)}\Bigr\}=0,
\label{eq:E17}
\end{eqnarray}
\begin{eqnarray}
\frac{1}{V}
\frac{\partial \Omega}{\partial {\bar \Phi}}
&=& \frac{T^4}{2}(-b_2(T)\Phi-b_3{\bar \Phi}^2+b_4{\bar \Phi}\Phi^2)
    \nonumber\\
&&  -6 N_{\rm f} T \int \frac{d^3 p}{(2\pi)^3}\Bigl\{
    \frac{e^{-2\beta E^-({\bf p})}}
         {g^{(-)}(\sigma,\Phi,{\bar \Phi};T,\mu)} 
    +\frac{e^{-\beta E^+({\bf p})}}
         {g^{(+)}(\sigma,\Phi,{\bar \Phi};T,\mu)}
    \Bigr\}=0,
\label{eq:E18}
\\
\frac{1}{V}
\frac{\partial \Omega}{\partial {\Phi}}
&=& \frac{T^4}{2}(-b_2(T){\bar \Phi}-b_3{\Phi}^2+b_4{\bar \Phi}^2\Phi)
    \nonumber\\
&&  -6 N_{\rm f} T \int \frac{d^3 p}{(2\pi)^3}\Bigl\{
    \frac{e^{-\beta E^-({\bf p})}}
         {g^{(-)}(\sigma,\Phi,{\bar \Phi};T,\mu)} 
    +\frac{e^{-2\beta E^+({\bf p})}}
         {g^{(+)}(\sigma,\Phi,{\bar \Phi};T,\mu)}
    \Bigr\}=0,
\label{eq:E19}
\end{eqnarray}
with 
\begin{eqnarray}
g^{(-)}(\sigma,\Phi,{\bar \Phi};T,\mu) 
&=& 1 + 3(\Phi+{\bar \Phi}e^{-\beta E^-({\bf p})})e^{-\beta E^-({\bf p})}
                                         + e^{-3\beta E^-({\bf p})}, \\
g^{(+)}(\sigma,\Phi,{\bar \Phi};T,\mu) 
&=& 1 + 3({\bar \Phi}+\Phi e^{-\beta E^+({\bf p})})e^{-\beta E^+({\bf p})}
                                         + e^{-3\beta E^+({\bf p})}, \\
h^{(-)}(\sigma,\Phi,{\bar \Phi};T,\mu) 
&=& \Phi e^{-\beta E^-({\bf p})}+2{\bar \Phi}e^{-2\beta E^-({\bf p})}
    +e^{-3\beta E^-({\bf p})}, \\  
h^{(+)}(\sigma,\Phi,{\bar \Phi};T,\mu)  
&=& {\bar \Phi} e^{-\beta E^+({\bf p})}+2{\Phi}e^{-2\beta E^+({\bf p})}
    +e^{-3\beta E^+({\bf p})},                            
\end{eqnarray}
where the effective coupling $G_{\rm s}^*$ is defined as 
$G_{\rm s}^* = 2G_{\rm s} + 12 G_{\rm s8} \sigma^2$.

Values of $\sigma$ are directly determined from minima of 
the real part of the thermodynamical potential, 
while $\Phi$, ${\bar \Phi}$ and $\rho_{\rm v}$ are obtained by solving Eqs. (\ref{eq:E16})-(\ref{eq:E19}).

Following Refs. \cite{Fukushima,Sasaki1}, 
we define the susceptibilities as 
\begin{eqnarray}
\chi 
&=& C^{-1} \nonumber \\
&=& \frac{1}{{\rm det} C}\left(
\begin{array}{cccc}
  C_{\Phi \Phi}C_{{\bar \Phi} {\bar \Phi}} - C_{\Phi {\bar \Phi}}C_{{\bar \Phi} \Phi}
& C_{\sigma {\bar \Phi}}C_{{\bar \Phi} \sigma} - C_{\sigma \Phi}C_{{\bar \Phi} {\bar \Phi}}  
& C_{\sigma \Phi} C_{\Phi {\bar \Phi}} - C_{\sigma {\bar \Phi}}C_{\Phi \Phi} \\
  C_{\Phi {\bar \Phi}}C_{{\bar \Phi} \sigma} - C_{\Phi \sigma}C_{{\bar \Phi} {\bar \Phi}}
& C_{\sigma \sigma}C_{{\bar \Phi} {\bar \Phi}} - C_{\sigma {\bar \Phi}}C_{{\bar \Phi} \sigma}  
& C_{\sigma {\bar \Phi}}C_{\Phi \sigma} - C_{\sigma \sigma}C_{\Phi {\bar \Phi}}\\
  C_{\Phi \sigma}C_{{\bar \Phi} \Phi} - C_{\Phi \Phi}C_{{\bar \Phi} \sigma}
& C_{\sigma \Phi}C_{{\bar \Phi} \sigma} - C_{\sigma \sigma}C_{{\bar \Phi} \Phi}
& C_{\sigma \sigma}C_{\Phi \Phi} - C_{\sigma \Phi}C_{\Phi \sigma}\\
\end{array}
\right),\\
\chi_{ij} &=& C^{-1}_{ij}~~~~(i,j=\sigma,\Phi,{\bar \Phi}),
\end{eqnarray}
where $C$ is the matrix of dimensionless curvatures 
\begin{eqnarray}
C &=& \left(
\begin{array}{cccc} 
\frac{\beta}{4G_{\rm s}^2 \Lambda}
\frac{\partial^2 \Omega}{\partial \sigma^2} 
& -\frac{\beta}{2G_{\rm s} \Lambda^2}
\frac{\partial^2 \Omega}{\partial \sigma \partial \Phi} 
& -\frac{\beta}{2G_{\rm s} \Lambda^2}
\frac{\partial^2 \Omega}{\partial \sigma \partial {\bar \Phi}} \\
-\frac{\beta}{2G_{\rm s} \Lambda^2}
\frac{\partial^2 \Omega}{\partial \Phi \partial \sigma} 
& \frac{\beta}{\Lambda^3}
\frac{\partial^2 \Omega}{\partial \Phi^2} 
& \frac{\beta}{\Lambda^3}
\frac{\partial^2 \Omega}{\partial \Phi \partial {\bar \Phi}} \\
-\frac{\beta}{2G_{\rm s} \Lambda^2}
\frac{\partial^2 \Omega}{\partial {\bar \Phi} \partial \sigma} 
& \frac{\beta}{\Lambda^3}
\frac{\partial^2 \Omega}{\partial {\bar \Phi} \partial \Phi} 
& \frac{\beta}{\Lambda^3}
\frac{\partial^2 \Omega}{\partial {\bar \Phi}^2} \\
\end{array}
\right).
\end{eqnarray}
In this study, the susceptibility is used to determine  
a pseudo-critical temperature of crossover.

\begin{figure}[htbp]
\begin{center}
 \includegraphics[width=7.5cm]{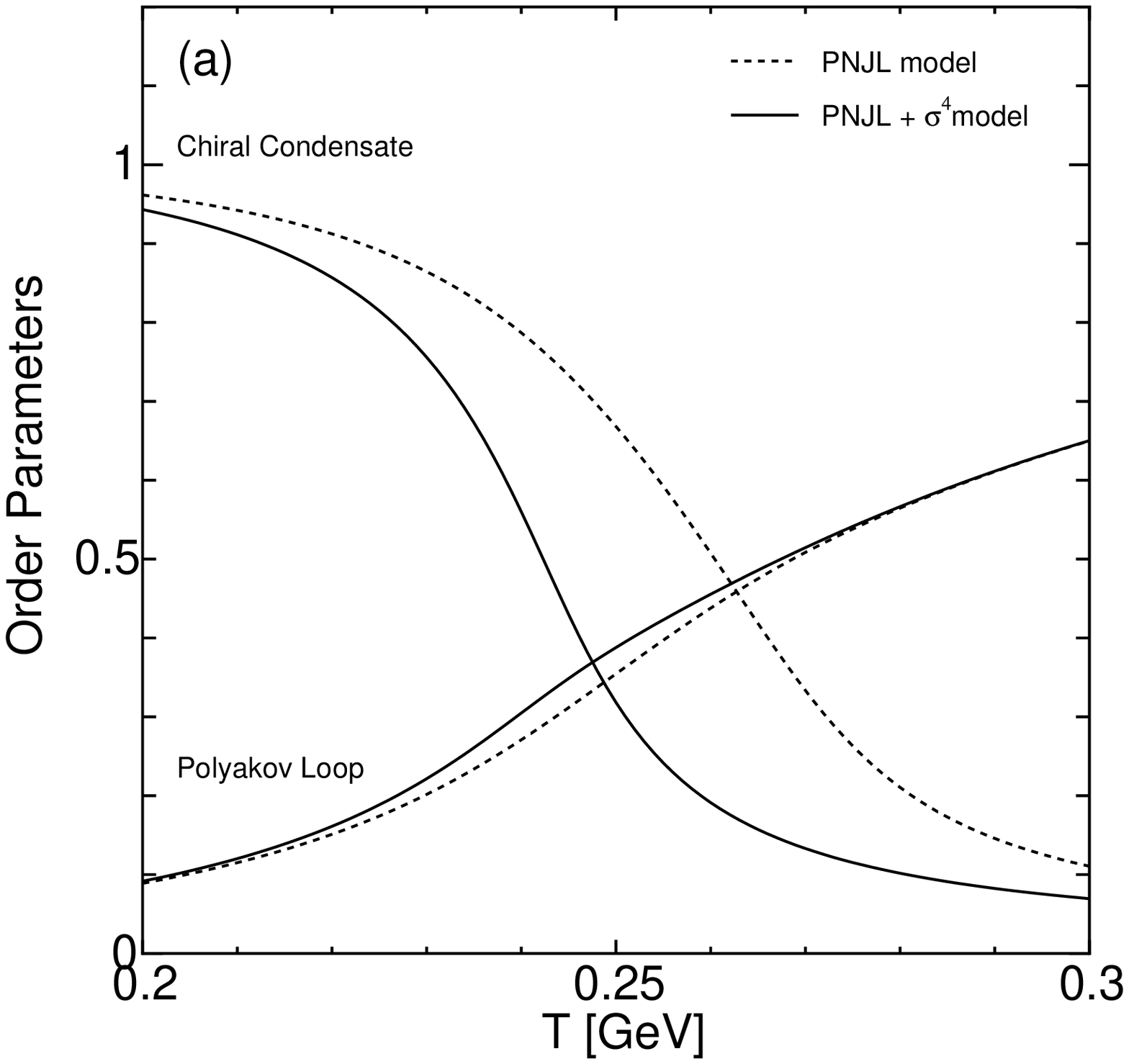} 
 \includegraphics[width=7.5cm]{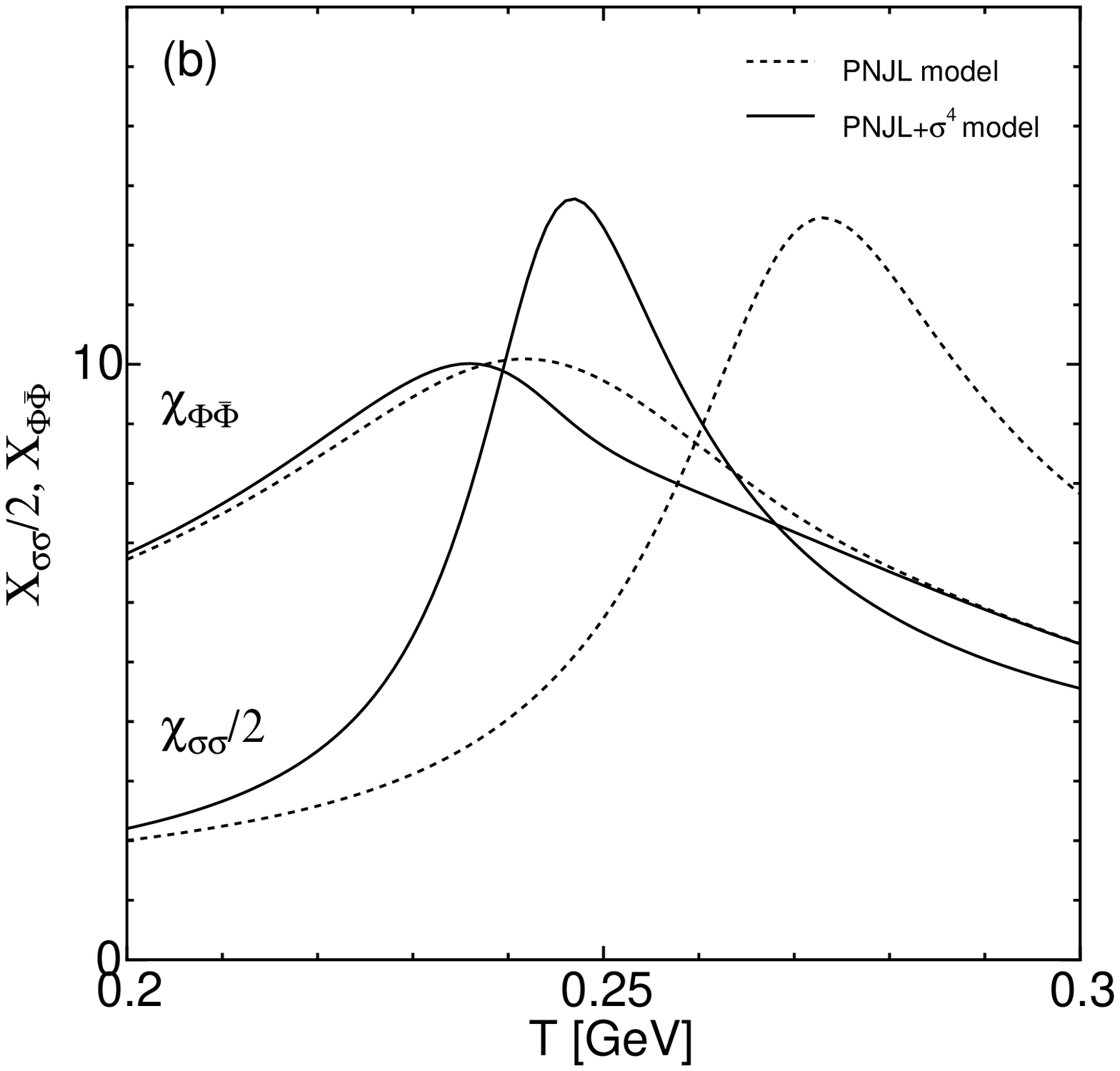} 
\end{center}
\caption{The left panel graphs the $T$ dependence of 
the chiral condensate 
$\langle \bar{q}q \rangle / \langle \bar{q}q \rangle_{T=\mu=0}$ and the Polyakov loop $\Phi$ 
at $\mu=0$.
The right panel graphs the chiral and Polyakov loop susceptibilities.}
\label{fig1}
\end{figure}
First we discuss the $T$ dependence of 
the chiral condensate, the Polyakov loop and their susceptibilities 
in the case of $\mu=0$. 
Figure 1 shows two results of the original PNJL and the PNJL$+\sigma^4$ models,  
since the vector type interaction does not contribute
to the phase transition when $\mu=0$. 
From the left panel, 
we can see that the $\sigma^4$  interaction makes the 
chiral phase transition sharper and its 
pseudo-critical temperature lower in the PNJL model as well as in the NJL model~\cite{Kashiwa1}. 
Similar effects are also seen in the three flavor NJL model~\cite{Osipov3,Osipov4}. 
Meanwhile, the $\sigma^4$ interaction affects the Polyakov loop little. 
As a result, the pseudo-critical temperature of the chiral
phase transition goes down to the vicinity of that of the deconfinement transition.

The pseudo-critical temperatures 
are calculated  also in  Refs.~\cite{Ratti1,Hansen}. 
Our regularization scheme is the same as that of Ref.~\cite{Hansen} but not 
as that of Ref.~\cite{Ratti1}, that is, in the present work 
the momentum cutoff is taken for both the vacuum and $T$-dependent terms 
in the square bracket of 
Eq.~(\ref{eq:E11}), while in Ref.~\cite{Ratti1} 
the cutoff is made only for the vacuum term. Consequently, 
our result is consistent with that of Ref.~\cite{Hansen}, 
but somewhat deviates from that of Ref.~\cite{Ratti1}.

The pseudo-critical temperatures ($T_{\rm c}$) can be clearly defined by the peak of the susceptibilities, as shown 
in the right panel of Fig.1.  
In the original PNJL model the pseudo-critical temperatures of the chiral and 
the deconfinement phase transitions 
are $T_{\rm c}=273$ MeV and 242 MeV, respectively. 
In the PNJL $+\sigma^4$ model, 
the corresponding values are $T_{\rm c}=247$ MeV and 236 MeV. 
The differences between the chiral and the deconfinement pseudo-critical temperatures are 31 MeV in the original PNJL 
model and 11 MeV in the PNJL$+\sigma^4$ model. Thus, the $\sigma^4$ interaction makes the difference smaller.

\begin{figure}[htbp]
\begin{center}
 \includegraphics[width=7.5cm]{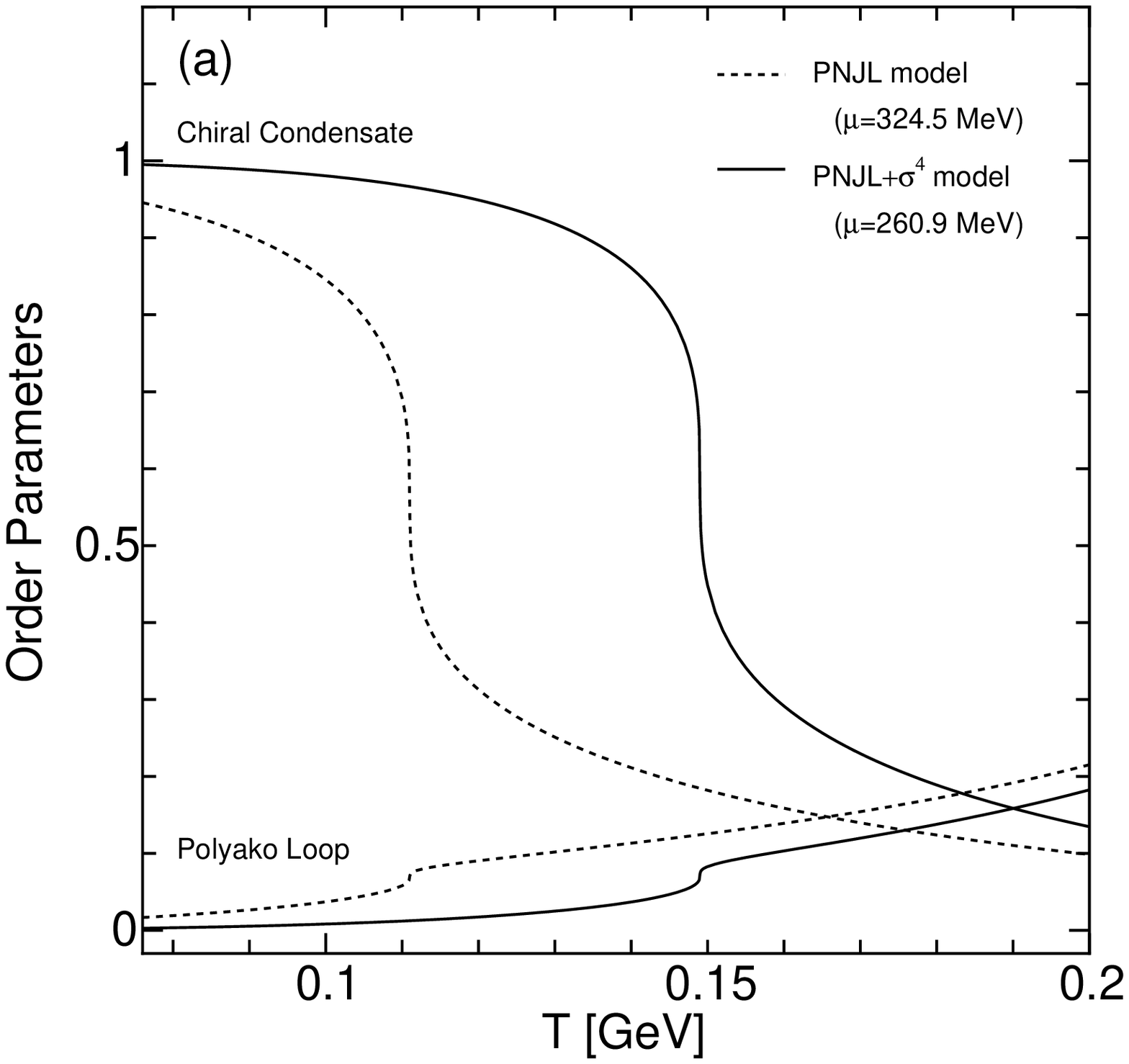} 
 \includegraphics[width=7.5cm]{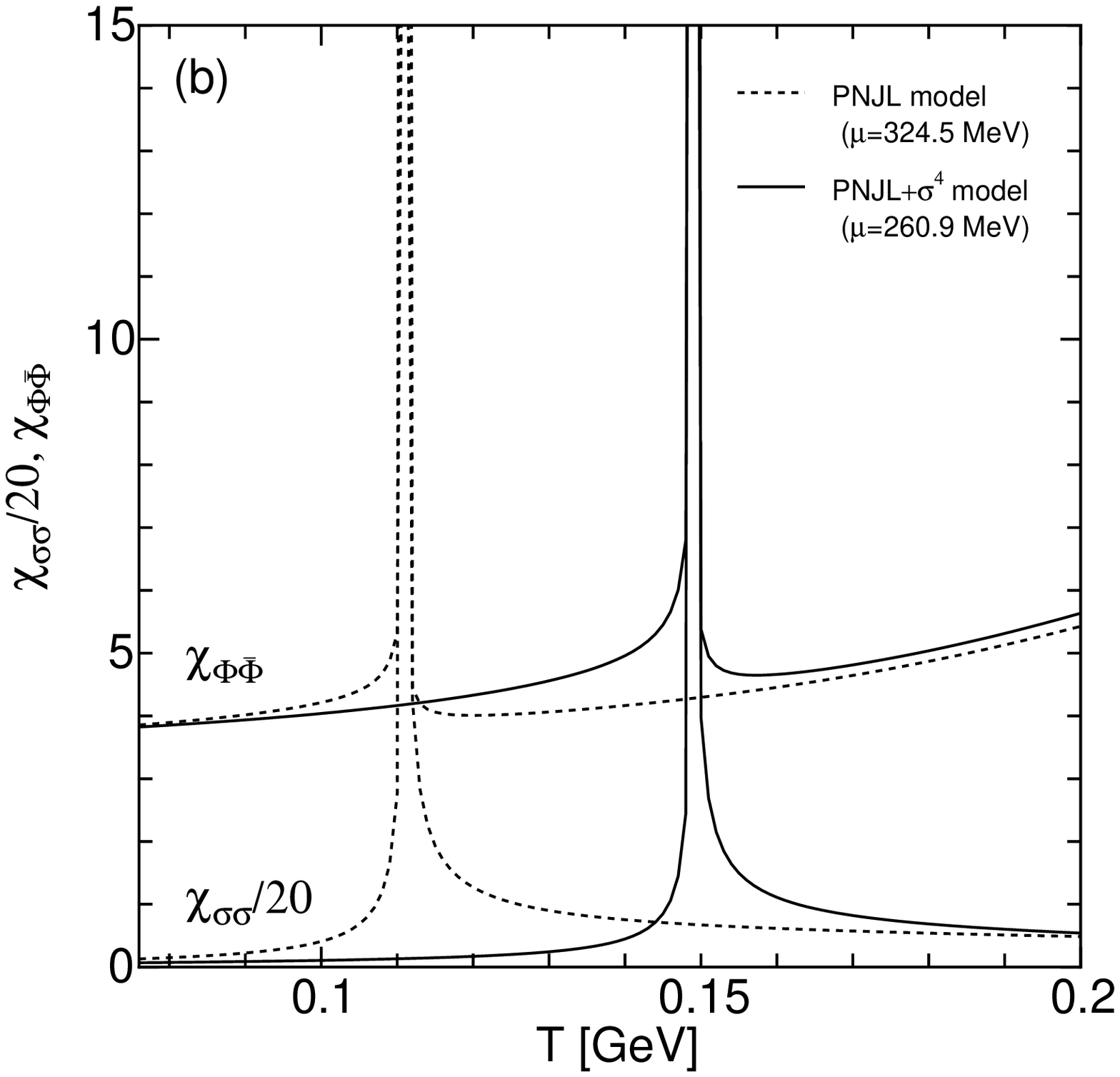} 
\end{center}
\caption{The left panel graphs the $T$ dependence of 
the chiral condensate $\langle \bar{q}q\rangle / \langle \bar{q}q \rangle_{T=\mu=0}$ and the Polyakov loop $\Phi$ 
near the CEP. 
The right panel graphs the chiral and Polyakov loop susceptibilities.}
\label{fig2}
\end{figure}
Second we discuss the behavior of the chiral condensate and the Polyakov loop near the CEP $(T_{\rm e},\mu_{\rm e})$. 
Figure 2 shows the $T$ dependence of 
the chiral condensate, the Polyakov loop and their susceptibilities near the CEP.  
As clearly seen in the right panel, both the chiral and Polyakov loop susceptibilities diverge at the CEP.
This indicates that the two phase transitions are second order at the CEP.
Further discussions on the CEP will be made in a forthcoming paper.

\begin{figure}[htbpt]
\begin{center}
 \includegraphics[width=7.5cm]{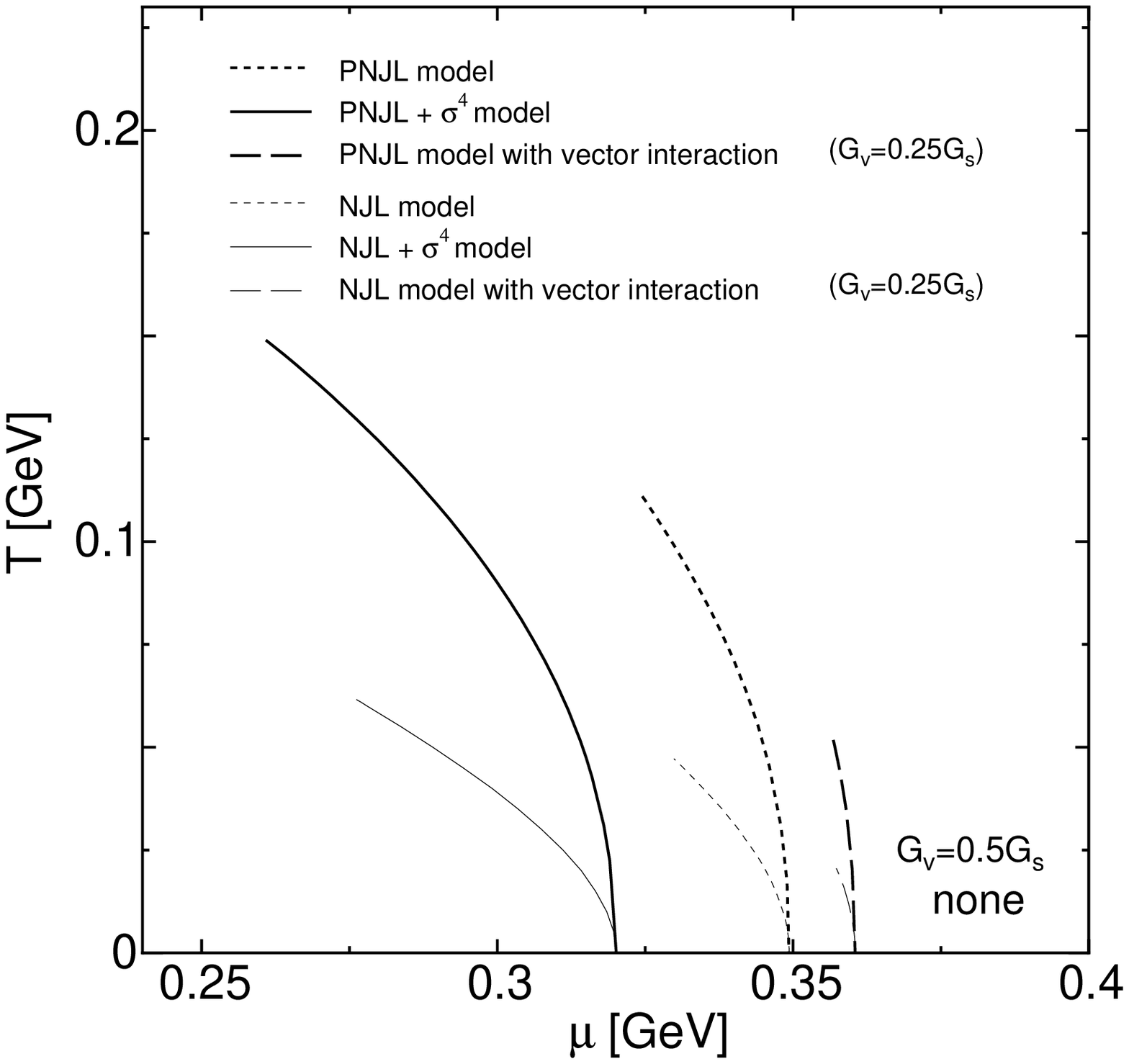}
 \includegraphics[width=7.3cm]{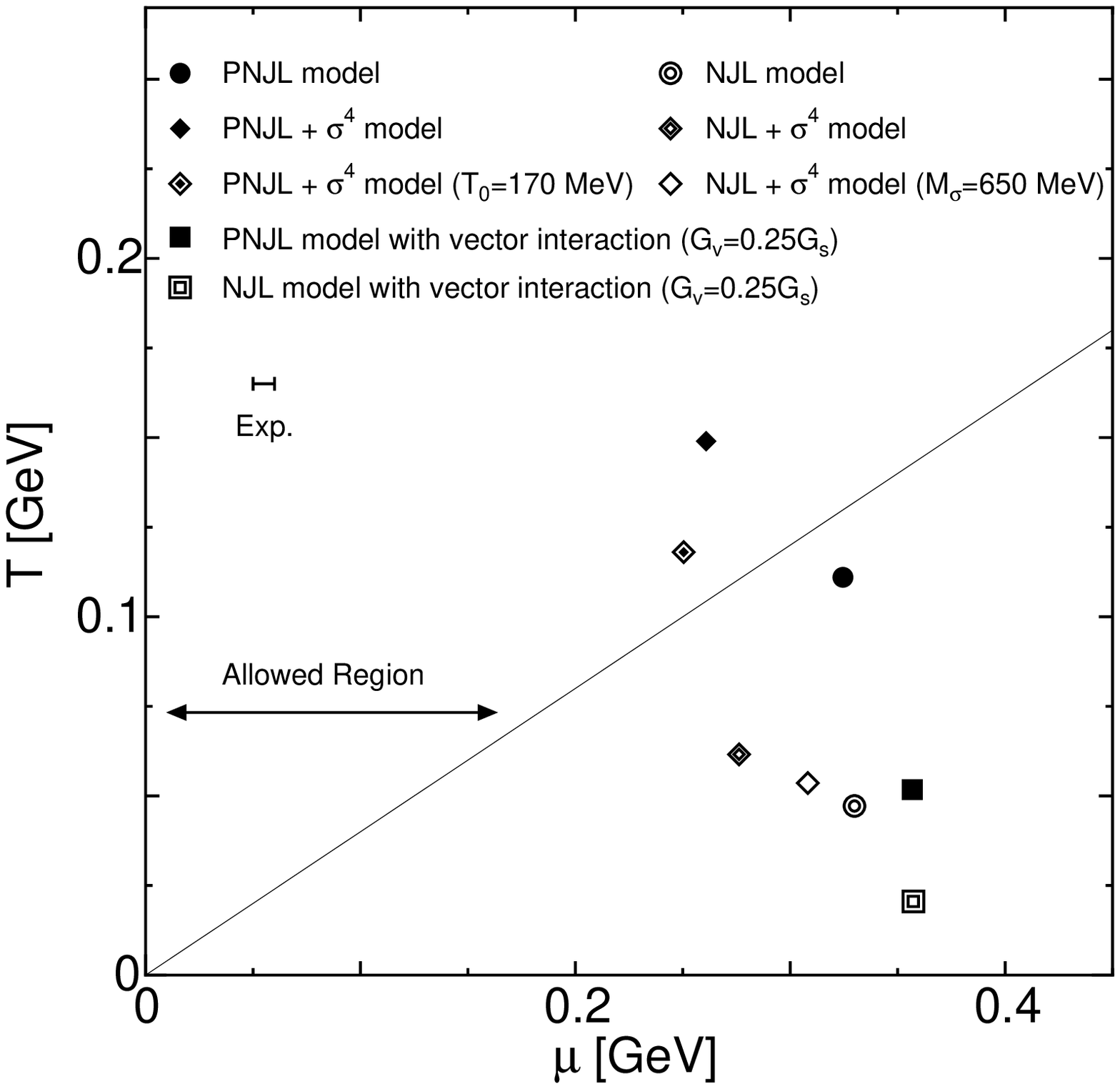} 
\end{center}
\caption{The left panel graphs the phase diagram on the $T-\mu$ plane.
The right panel shows positions of the CEP in  
several versions of the NJL model and the PNJL model. The allowed region of the $\mu_{\rm e}/T_{\rm e}$ suggested by 
a lattice simulation is also indicated.
The values of $T_{\rm e}$ and $\mu_{\rm e}$ in several models are explicitly
given in Table III.}
\label{fig3}
\end{figure}
Finally we show the phase diagram in Fig. 3. 
In the region of $T<T_{\rm e}$ and $\mu>\mu_{\rm e}$  
the chiral and deconfinement transitions are first order and occur at the same time. 
This can be understood by the generalized Clausius-Clapeyron relation 
for systems with multiple order parameters, which ensures that their 
discontinuities appear at the same $T$ and $\mu$ when all the transitions are first order~\cite{Barducci}. 
Thus, the deconfinement phase transition seems to be dragged by the chiral phase transition.

It is known~\cite{Sasaki1} that the Polyakov loop susceptibility 
has a broad bump in the larger 
$T$ region in addition to 
a sharp peak in the smaller $T$ region. 
In Ref.~\cite{Sasaki1}, the sharp peak was interpreted 
as a reflection of the chiral phase transition and 
the broad bump was then identified 
with the critical temperature for the deconfinement phase transition. 
The broad bump does not disappear even if the $\sigma^4$ interaction 
is introduced. 
 However, 
we simply define the critical temperature with the sharp peak. 
The critical temperature thus defined for the deconfinement phase transition 
approaches that for the chiral phase transition as $\mu$ increases to 
$\mu_{\rm e}$, and both the critical temperatures agree with each other when 
$\mu \ge \mu_{\rm e}$. This behavior is consistent with 
the generalized Clausius-Clapeyron relation 
among multiple first-order phase transitions~\cite{Barducci}.

The left panel of Fig. 3 shows phase transition lines of six models. 
Comparing the original PNJL and the PNJL+$\sigma^4$ models, we see that  
the $\sigma^4$  interaction makes the CEP move toward higher $T$ and lower $\mu$.
Meanwhile, comparison between the original PNJL model and the PNJL model with vector interaction shows that 
the vector type interaction makes the first-order phase transition weak  
and does the CEP move toward lower $T$ and higher $\mu$, that is, in the direction opposite to the case of 
the $\sigma^4$ interaction. The CEP in the PNJL model is always located at $T$ higher than that in the NJL model, 
even if either the vector or the $\sigma^4$ interaction are added to the models.

The right panel of Fig. 3 shows positions of the CEP in several versions of the NJL and the PNJL models. 
Results of all models are far from the empirical value, but 
the result of the PNJL+$\sigma^4$ model is closest. 
It is suggested in the recent lattice analysis~\cite{Ejiri} that the possible region where the CEP exists is 
$\mu_{\rm e}/T_{\rm e} \sik 2.5$. 
The allowed region is denoted by the bidirectional arrow in the right panel. 
Only the PNJL+$\sigma^4$ model satisfies this restriction. 
So far we took $T_0$=270 MeV, but the pseudo-critical temperature $T_{\rm c}$ evaluated by adopting this value of 
$T_0$ is somewhat higher than the prediction of a lattice QCD calculation~\cite{Karsch}.
So, we have rescaled $T_0$ so that the average of the chiral and deconfinement pseudo-critical 
temperatures can agree with that of the lattice QCD at $\mu=0$. The rescaled $T_0$ is 170 MeV. 
The CEP given by the PNJL+$\sigma^4$ model yields $\mu_{\rm e}/T_{\rm e} \sim 1.8$ when $T_0$=270 MeV and $2.1$ when 
$T_0$=170 MeV. 
Both the results satisfy the restriction $\mu_{\rm e}/T_{\rm e} \sik 2.5$. 
Moreover, the latest lattice analysis~\cite{Ejiri} points out $T_{\rm e}/T_{\rm c} \sim 0.8$. The ratio calculated 
with the PNJL+$\sigma^4$ model 
is 0.6-0.7 
in both cases of $T_0$=170 and 270 MeV and consistent with the prediction of the lattice analysis~\cite{Ejiri}. 

\begin{table}[b]
\begin{center}
\begin{tabular}{llll}
\hline
~~~~model~~~~ & $T_{\rm e}$ [MeV]~~~~ & $\mu_{\rm e}$ [MeV]~~~~ & $\mu_{\rm e}/T_{\rm e}$~~~~ 
\\
\hline
~~~~NJL
& 0.047~~~~ 
& 0.330~~~~
&   7.0~~~~
\\
~~~~NJL$+\sigma^4$
& 0.062
& 0.276
&   4.5
\\
~~~~NJL$+\sigma^4$ ($M_\sigma=650$ MeV)
& 0.054
& 0.308
&   5.7
\\  
~~~~NJL with vector interaction ($G_{\rm v}=0.25 G_{\rm s}$)
& 0.021
& 0.357
&   17
\\  
~~~~PNJL
& 0.111
& 0.325
&   2.9
\\ 
~~~~PNJL$+\sigma^4$
& 0.149
& 0.261
&   1.8
\\  
~~~~PNJL$+\sigma^4$ ($T_0=170$ MeV)
& 0.118 
& 0.250 
&   2.1
\\  
~~~~PNJL with vector interaction ($G_{\rm v}=0.25 G_{\rm s}$)~~~~
& 0.052
& 0.357
&   6.9
\\  
\hline
\end{tabular}
\caption{The values of $T_{\rm e}$, $\mu_{\rm e}$ and $\mu_{\rm e}/T_{\rm e}$ in several models.}
\end{center}
\end{table}

In summary, we have investigated effects of the $\sigma^4$  and the vector interactions on the position of the CEP 
and the interplay between the chiral and deconfinement phase transitions. 
In the case of $\mu=0$, the $\sigma^4$ 
interaction shifts the pseudo-critical temperature of the chiral transition
to the vicinity of that of the deconfinement transition. 
As for the CEP, the $\sigma^4$ interaction shifts it largely toward higher $T$ and 
lower $\mu$, while the vector type interaction shifts it in the opposite direction. 
The CEP calculated with the PNJL+$\sigma^4$ model is closest to the
empirical one and is in good agreement with the restriction $\mu_{\rm e}/T_{\rm e} \sik 2.5$ given by the recent 
lattice analysis. 
Thus, it is quite interesting to investigate roles 
of the $\sigma^4$ interaction in 
other thermodynamic quantities such as pressure and 
quark number density.

\noindent
\begin{acknowledgments}
The authors thank Prof. M. Tachibana for useful discussions. 
H.K. also thanks Prof. T. Kunihiro, Prof. M. Imachi and Prof. H. Yoneyama for useful discussions. 
This work has been supported in part by the Grants-in-Aid for Scientific Research 
(18540280) of Education, Science, Sports, and Culture of Japan.
\end{acknowledgments}



\begin{thebibliography}{19}
\expandafter\ifx\csname natexlab\endcsname\relax\def\natexlab#1{#1}\fi
\expandafter\ifx\csname bibnamefont\endcsname\relax
  \def\bibnamefont#1{#1}\fi
\expandafter\ifx\csname bibfnamefont\endcsname\relax
  \def\bibfnamefont#1{#1}\fi
\expandafter\ifx\csname citenamefont\endcsname\relax
  \def\citenamefont#1{#1}\fi
\expandafter\ifx\csname url\endcsname\relax
  \def\url#1{\texttt{#1}}\fi
\expandafter\ifx\csname urlprefix\endcsname\relax\def\urlprefix{URL }\fi
\providecommand{\bibinfo}[2]{#2}
\providecommand{\eprint}[2][]{\url{#2}}

%

\bibitem[{\citenamefont{Kogut et~al.}(1983)\citenamefont{Kogut, Stone, {H. W.
  Wyld}, {W. R. Gibbs}, Shigemitsu, {S. H. Shenker}, and {D. K.
  Sinclair}}}]{Kog}
\bibinfo{author}{\bibfnamefont{J.}~\bibnamefont{Kogut}},
  \bibinfo{author}{\bibfnamefont{M.}~\bibnamefont{Stone}},
  \bibinfo{author}{\bibnamefont{{H. W. Wyld}}},
  \bibinfo{author}{\bibnamefont{{W. R. Gibbs}}},
  \bibinfo{author}{\bibfnamefont{J.}~\bibnamefont{Shigemitsu}},
  \bibinfo{author}{\bibnamefont{{S. H. Shenker}}}, \bibnamefont{and}
  \bibinfo{author}{\bibnamefont{{D. K. Sinclair}}}, \bibinfo{journal}{Phys.\
  Rev.\ Lett.} \textbf{\bibinfo{volume}{50}}, \bibinfo{pages}{393}
  (\bibinfo{year}{1983}).

\bibitem[{\citenamefont{Fodor and Katz}(2002)}]{ZF}
\bibinfo{author}{\bibfnamefont{Z.}~\bibnamefont{Fodor}} \bibnamefont{and}
\bibinfo{author}{\bibfnamefont{S.}~\bibfnamefont{D.}~\bibnamefont{Katz}},  
\bibinfo{journal}{J. High Energy Phys.} \textbf{\bibinfo{volume}{03}},
\bibinfo{pages}{014} (\bibinfo{year}{2002});
\bibinfo{journal}{Prog. Theor. Phys. Suppl.} \textbf{\bibinfo{volume}{153}},
\bibinfo{pages}{86} (\bibinfo{year}{2004}).

\bibitem[{\citenamefont{Ejiri}(2007)}]{Ejiri}
\bibinfo{author}{\bibfnamefont{S.}~\bibnamefont{Ejiri}},
  \bibinfo{howpublished}{arXiv:hep-lat/0706.3549} (\bibinfo{year}{2007});
\bibinfo{howpublished}{arXiv:hep-lat/0710.0653} (\bibinfo{year}{2007}).
 
\bibitem[{\citenamefont{Nambu and Jona-Lasinio}(1961{\natexlab{a}})}]{NJ1}
\bibinfo{author}{\bibfnamefont{Y.}~\bibnamefont{Nambu}} \bibnamefont{and}
  \bibinfo{author}{\bibfnamefont{G.}~\bibnamefont{Jona-Lasinio}},
  \bibinfo{journal}{Phys.\ Rev.} \textbf{\bibinfo{volume}{122}},
  \bibinfo{pages}{345} (\bibinfo{year}{1961}); 
  \bibinfo{journal}{Phys.\ Rev.} \textbf{\bibinfo{volume}{124}},
  \bibinfo{pages}{246} (\bibinfo{year}{1961}).

\bibitem[{\citenamefont{Asakawa and Yazaki}(1989)}]{AY}
\bibinfo{author}{\bibfnamefont{M.}~\bibnamefont{Asakawa}} \bibnamefont{and}
  \bibinfo{author}{\bibfnamefont{K.}~\bibnamefont{Yazaki}},
  \bibinfo{journal}{Nucl.\ Phys.} \textbf{\bibinfo{volume}{A504}},
  \bibinfo{pages}{668} (\bibinfo{year}{1989}). 

\bibitem[{\citenamefont{Scavenius et~al.}(2000)\citenamefont{Scavenius, {{\'
  A}. M{\' o}csy}, {I. N. Mishustin}, and {D. H. Rischke}}}]{Sca}
\bibinfo{author}{\bibfnamefont{O.}~\bibnamefont{Scavenius}},
  \bibinfo{author}{\bibnamefont{{{\' A}. M{\' o}csy}}},
  \bibinfo{author}{\bibnamefont{{I. N. Mishustin}}}, \bibnamefont{and}
  \bibinfo{author}{\bibnamefont{{D. H. Rischke}}}, \bibinfo{journal}{Phys.\
  Rev.\ C} \textbf{\bibinfo{volume}{64}}, \bibinfo{pages}{045202}
  (\bibinfo{year}{2001}).

\bibitem[{\citenamefont{Kiriyama et al.}(2001)}]{KMT}
\bibinfo{author}{\bibfnamefont{O.}~\bibnamefont{Kiriyama}},
  \bibinfo{author}{\bibfnamefont{M.}~\bibnamefont{Maruyama}},
  \bibnamefont{and}
  \bibinfo{author}{\bibfnamefont{F.}~\bibfnamefont{Takagi}},
  \bibinfo{journal}{Phys. Rev. \ D} \textbf{\bibinfo{volume}{63}},
  \bibinfo{pages}{116009} (\bibinfo{year}{2001}).

\bibitem[{\citenamefont{Hashimot et al.}(2005)}]{HTF}
\bibinfo{author}{\bibfnamefont{Y.}~\bibnamefont{Hashimoto}},
  \bibinfo{author}{\bibfnamefont{Y.}~\bibnamefont{Tsue}},
  \bibnamefont{and}
  \bibinfo{author}{\bibfnamefont{H.}~\bibfnamefont{Fujii}},
  \bibinfo{journal}{Prog. Theor. Phys. } \textbf{\bibinfo{volume}{114}},
  \bibinfo{pages}{595} (\bibinfo{year}{2005}).

\bibitem[{\citenamefont{Lacey et al.}(2007)}]{Lacey}
\bibinfo{author}{\bibfnamefont{R.}~\bibfnamefont{A.}~\bibnamefont{Lacey}},
\bibinfo{author}{\bibfnamefont{N.}~\bibfnamefont{N.}~\bibnamefont{Ajitanand}},
\bibinfo{author}{\bibfnamefont{J.}~\bibfnamefont{M.}~\bibnamefont{Alexander}},
\bibinfo{author}{\bibfnamefont{P.}~\bibnamefont{Chung}},
\bibinfo{author}{\bibfnamefont{J.}~\bibnamefont{Jia}},
\bibinfo{author}{\bibfnamefont{A.}~\bibnamefont{Taranenko}}, 
\bibnamefont{and}
\bibinfo{author}{\bibfnamefont{P.}~\bibnamefont{Danielewicz}}, 
  \bibinfo{howpublished}{arXiv:nucl-ex/0708.3512} (\bibinfo{year}{2007}).   

\bibitem[{\citenamefont{Kashiwa et al}(2006)}]{Kashiwa1}
\bibinfo{author}{\bibfnamefont{K.}~\bibnamefont{Kashiwa}}, 
\bibinfo{author}{\bibfnamefont{H.}~\bibnamefont{Kouno}}, 
\bibinfo{author}{\bibfnamefont{T.}~\bibnamefont{Sakaguchi}}, 
\bibinfo{author}{\bibfnamefont{M.}~\bibnamefont{Matsuzaki}}, 
\bibnamefont{and}
\bibinfo{author}{\bibfnamefont{M.}~\bibnamefont{Yahiro}},
\bibinfo{journal}{Phys. Lett.\ B} \textbf{\bibinfo{volume}{647}},
\bibinfo{pages}{446} (\bibinfo{year}{2007}). 

\bibitem[{\citenamefont{Zschiesche et al.}(2005)}]{ZZSS}
\bibinfo{author}{\bibfnamefont{D.}~\bibnamefont{Zschiesche}},
\bibinfo{author}{\bibfnamefont{G.}~\bibnamefont{Zeeb}},
\bibinfo{author}{\bibfnamefont{S.}~\bibnamefont{Schramm}}
\bibnamefont{and}
\bibinfo{author}{\bibfnamefont{H.}~\bibnamefont{St\"{o}cker}},
\bibinfo{journal}{J. Phys. G: Nucl. Part. Phys. } \textbf{\bibinfo{volume}{31}},
\bibinfo{pages}{935} (\bibinfo{year}{2005}).

\bibitem[{\citenamefont{Zschiesche et al.}(2005)}]{ZZS}
\bibinfo{author}{\bibfnamefont{D.}~\bibnamefont{Zschiesche}},
\bibinfo{author}{\bibfnamefont{G.}~\bibnamefont{Zeeb}},
\bibnamefont{and}
\bibinfo{author}{\bibfnamefont{S.}~\bibnamefont{Schramm}},
  \bibinfo{howpublished}{arXiv:nucl-th/0602073} (\bibinfo{year}{2006}). 
  
\bibitem[{\citenamefont{Meisinger et al.}(2007)}]{Meisinger}
\bibinfo{author}{\bibfnamefont{P.}~\bibnamefont{N.}}~\bibnamefont{Meisinger},
\bibnamefont{and}
\bibinfo{author}{\bibfnamefont{M.}~\bibnamefont{C.}}~\bibnamefont{Ogilvie},  
  \bibinfo{journal}{Phys. Lett.\ B} \textbf{\bibinfo{volume}{379}},
  \bibinfo{pages}{163} (\bibinfo{year}{1996}). 

\bibitem[{\citenamefont{Fukushima}(1989)}]{Fukushima}
\bibinfo{author}{\bibfnamefont{K.}~\bibnamefont{Fukushima}}, 
  \bibinfo{journal}{Phys. Lett.\ B} \textbf{\bibinfo{volume}{591}},
  \bibinfo{pages}{277} (\bibinfo{year}{2004}). 

\bibitem[{\citenamefont{{S. K. Ghosh} et al.}(2006)}]{Ghos}
\bibinfo{author}{\bibnamefont{{S. K. Ghosh}}},
  \bibinfo{author}{\bibnamefont{{T. K. Mukherjee}}},
  \bibinfo{author}{\bibnamefont{{M. G. Mustafa}}}, \bibnamefont{and}
  \bibinfo{author}{\bibfnamefont{R.}~\bibnamefont{Ray}},
  \bibinfo{journal}{Phys.\ Rev.\ D} \textbf{\bibinfo{volume}{73}},
  \bibinfo{pages}{114007} (\bibinfo{year}{2006}).
					    
\bibitem[{\citenamefont{Megias et al.}(2006)}]{Megias}
\bibinfo{author}{\bibfnamefont{E.}~\bibnamefont{Meg{$\acute{\i}$}as}},
\bibinfo{author}{\bibfnamefont{E.}~\bibnamefont{R.}~\bibnamefont{Arriola}},
\bibnamefont{and}
\bibinfo{author}{\bibfnamefont{L.}~\bibnamefont{L.}~\bibnamefont{Salcedo}},  
  \bibinfo{journal}{Phys. Rev.\ D} \textbf{\bibinfo{volume}{74}},
  \bibinfo{pages}{065005} (\bibinfo{year}{2006}). 

\bibitem[{\citenamefont{Ratti et al.}(2006)}]{Ratti1}
\bibinfo{author}{\bibfnamefont{C.}~\bibnamefont{Ratti}},
\bibinfo{author}{\bibfnamefont{M.}~\bibfnamefont{A.}~\bibnamefont{Thaler}},
\bibnamefont{and}
\bibinfo{author}{\bibfnamefont{W.}~\bibnamefont{Weise}},  
  \bibinfo{journal}{Phys. Rev.\ D} \textbf{\bibinfo{volume}{73}},
  \bibinfo{pages}{014019} (\bibinfo{year}{2006}). 

\bibitem[{\citenamefont{Ratti et al.}(2007)}]{Ratti2}
\bibinfo{author}{\bibfnamefont{C.}~\bibnamefont{Ratti}},
\bibinfo{author}{\bibfnamefont{S.}~\bibnamefont{R\"{o}{\ss}ner}},
\bibinfo{author}{\bibfnamefont{M.}~\bibfnamefont{A.}~\bibnamefont{Thaler}},
\bibnamefont{and}
\bibinfo{author}{\bibfnamefont{W.}~\bibnamefont{Weise}},  
  \bibinfo{journal}{Eur. Phys. J.\ C} \textbf{\bibinfo{volume}{49}},
  \bibinfo{pages}{213} (\bibinfo{year}{2007}).

\bibitem[{\citenamefont{Rossner et al.}(2007)}]{Rossner}
\bibinfo{author}{\bibfnamefont{S.}~\bibnamefont{R\"{o}{\ss}ner}},
\bibinfo{author}{\bibfnamefont{C.}~\bibnamefont{Ratti}},
\bibnamefont{and}
\bibinfo{author}{\bibfnamefont{W.}~\bibnamefont{Weise}},  
  \bibinfo{journal}{Phys. Rev.\ D} \textbf{\bibinfo{volume}{75}},
  \bibinfo{pages}{034007} (\bibinfo{year}{2007}). 

\bibitem[{\citenamefont{Hansen et al.}(2007)}]{Hansen}
\bibinfo{author}{\bibfnamefont{H.}~\bibnamefont{Hansen}}, 
\bibinfo{author}{\bibfnamefont{W.}~\bibfnamefont{M.}~\bibnamefont{Alberico}},
\bibinfo{author}{\bibfnamefont{A.}~\bibnamefont{Beraudo}}, 
\bibinfo{author}{\bibfnamefont{A.}~\bibnamefont{Molinari}},
\bibinfo{author}{\bibfnamefont{M.}~\bibnamefont{Nardi}},
\bibnamefont{and}
\bibinfo{author}{\bibfnamefont{C.}~\bibnamefont{Ratti}}, 
  \bibinfo{journal}{Phys. Rev.\ D} \textbf{\bibinfo{volume}{75}},
  \bibinfo{pages}{065004} (\bibinfo{year}{2007}). 

\bibitem[{\citenamefont{Sasaki et al.}(2007)}]{Sasaki1}
\bibinfo{author}{\bibfnamefont{C.}~\bibnamefont{Sasaki}},
\bibinfo{author}{\bibfnamefont{B.}~\bibnamefont{Friman}},
\bibnamefont{and}
\bibinfo{author}{\bibfnamefont{K.}~\bibnamefont{Redlich}}, 
\bibinfo{journal}{Phys. Rev.\ D} \textbf{\bibinfo{volume}{75}},
  \bibinfo{pages}{074013} (\bibinfo{year}{2007}). 
  
\bibitem[{\citenamefont{Schaefer}(2007)}]{Schaefer}
\bibinfo{author}{\bibfnamefont{B. -J.}~\bibnamefont{Schaefer}},
\bibinfo{author}{\bibfnamefont{J. M.}~\bibnamefont{Pawlowski}},
\bibnamefont{and}
\bibinfo{author}{\bibfnamefont{J.}~\bibnamefont{Wambach}},
  \bibinfo{journal}{Phys. Rev.\ D} \textbf{\bibinfo{volume}{76}},
  \bibinfo{pages}{074023} (\bibinfo{year}{2007}). 

\bibitem[{\citenamefont{Buballa}(1996)}]{Buballa}
\bibinfo{author}{\bibfnamefont{M.}~\bibnamefont{Buballa}},
  \bibinfo{journal}{Nucl.\ Phys.} \textbf{\bibinfo{volume}{A611}},
  \bibinfo{pages}{393} (\bibinfo{year}{1996}). 

\bibitem[{\citenamefont{Kitazawa et al. }(2002)}]{KKKN}
\bibinfo{author}{\bibfnamefont{M.}~\bibnamefont{Kitazawa}},
{\bibfnamefont{T.}~\bibnamefont{Koide}},
{\bibfnamefont{T.}~\bibnamefont{Kunihiro}},
\bibnamefont{and}
\bibinfo{author}{\bibfnamefont{Y.}~\bibnamefont{Nemoto}},
\bibinfo{journal}{Prog. Theor. Phys.} \textbf{\bibinfo{volume}{108}},
\bibinfo{pages}{929} (\bibinfo{year}{2002}). 

\bibitem[{\citenamefont{Kashiwa et al}(2006)}]{Kashiwa2}
\bibinfo{author}{\bibfnamefont{K.}~\bibnamefont{Kashiwa}}, 
\bibinfo{author}{\bibfnamefont{M.}~\bibnamefont{Matsuzaki}}, 
\bibinfo{author}{\bibfnamefont{H.}~\bibnamefont{Kouno}}, 
\bibnamefont{and}
\bibinfo{author}{\bibfnamefont{M.}~\bibnamefont{Yahiro}},
\bibinfo{journal}{Phys. Lett.\ B} \textbf{\bibinfo{volume}{657}},
\bibinfo{pages}{143} (\bibinfo{year}{2007}).

\bibitem[{\citenamefont{Blaschke}(2007)}]{Blaschke}
\bibinfo{author}{\bibfnamefont{D.}~\bibnamefont{Blaschke}},
\bibinfo{author}{\bibfnamefont{T.}~\bibnamefont{Kl\"{a}hn}},
\bibnamefont{and}
\bibinfo{author}{\bibfnamefont{F.}~\bibnamefont{Sandin}},
  \bibinfo{howpublished}{arXiv:nucl-th/0708.4216} (\bibinfo{year}{2007}). 

\bibitem[{\citenamefont{Osipov et al. }(2006)}]{Osipov1}
\bibinfo{author}{\bibfnamefont{A.}~\bibfnamefont{A.}~\bibnamefont{Osipov}},
{\bibfnamefont{B.}~\bibnamefont{Hiller}},
\bibnamefont{and} 
{\bibfnamefont{J.}~\bibnamefont{da Provid\^encia}},
\bibinfo{journal}{Phys. Lett.\ B} \textbf{\bibinfo{volume}{634}},
\bibinfo{pages}{48} (\bibinfo{year}{2006}). 

\bibitem[{\citenamefont{Osipov et al. }(2006)}]{Osipov2}
\bibinfo{author}{\bibfnamefont{A.}~\bibfnamefont{A.}~\bibnamefont{Osipov}},
{\bibfnamefont{B.}~\bibnamefont{Hiller}}, 
{\bibfnamefont{J.}~\bibnamefont{Moreira}}, 
\bibnamefont{and} 
{\bibfnamefont{A.}~\bibfnamefont{H.}~\bibnamefont{Blin}},
\bibinfo{journal}{Eur. Phys. J.\ C} \textbf{\bibinfo{volume}{46}},
\bibinfo{pages}{225} (\bibinfo{year}{2006}). 

\bibitem[{\citenamefont{Osipov et al. }(2006)}]{Osipov3}
\bibinfo{author}{\bibfnamefont{A.}~\bibfnamefont{A.}~\bibnamefont{Osipov}},
{\bibfnamefont{B.}~\bibnamefont{Hiller}},
{\bibfnamefont{J.}~\bibnamefont{Moreira}},
{\bibfnamefont{A.}~\bibnamefont{H.}~\bibnamefont~{Blin}},
\bibnamefont{and} 
{\bibfnamefont{J.}~\bibnamefont{da Provid\^encia}},
\bibinfo{journal}{Phys. Lett.\ B} \textbf{\bibinfo{volume}{646}},
\bibinfo{pages}{91} (\bibinfo{year}{2007}). 

\bibitem[{\citenamefont{Osipov et al.}(2007)}]{Osipov4}
\bibinfo{author}{\bibfnamefont{A.}~\bibfnamefont{A.}~\bibnamefont{Osipov}},
\bibinfo{author}{\bibfnamefont{B.}~\bibnamefont{Hiller}},
\bibinfo{author}{\bibfnamefont{J.}~\bibnamefont{Moreira}},
\bibnamefont{and}
\bibinfo{author}{\bibfnamefont{A.}~\bibfnamefont{H.}~\bibnamefont{Blin}},  
  \bibinfo{howpublished}{arXiv:hep-ph/0709.3507} (\bibinfo{year}{2007}). 
  
\bibitem[{\citenamefont{Kapusta}(1989)}]{Kapusta}
\bibinfo{author}{\bibfnamefont{J. I.}~\bibnamefont{Kapusta}},  
 \bibinfo{howpublished}{{\it Finite-temperature field theory} 
(Cambridge University Press, 1989)}. 
  
\bibitem[{\citenamefont{Bellac}(1996)}]{Bellac}
\bibinfo{author}{\bibfnamefont{M. Le}~\bibnamefont{Bellac}},
\bibinfo{howpublished}{{\it Thermal Field Theory}
(Cambridge University Press, 1996)}.

\bibitem[{\citenamefont{Boyd et al.}(1996)}]{Boyd}
\bibinfo{author}{\bibfnamefont{G.}~\bibnamefont{Boyd}},
\bibinfo{author}{\bibfnamefont{J.}~\bibnamefont{Engels}},
\bibinfo{author}{\bibfnamefont{F.}~\bibnamefont{Karsch}},
\bibinfo{author}{\bibfnamefont{E.}~\bibnamefont{Laermann}},
\bibinfo{author}{\bibfnamefont{C.}~\bibnamefont{Legeland}},
\bibinfo{author}{\bibfnamefont{M.}~\bibnamefont{L\"{u}tgemeier}},
\bibnamefont{and}
\bibinfo{author}{\bibfnamefont{B.}~\bibnamefont{Petersson}},
 \bibinfo{journal}{Nucl. Phys.} \textbf{\bibinfo{volume}{B469}},
\bibinfo{pages}{419} (\bibinfo{year}{1996}). 

\bibitem[{\citenamefont{Kaczmarek}(2002)}]{Kaczmarek}
\bibinfo{author}{\bibfnamefont{O.}~\bibnamefont{Kaczmarek}},
\bibinfo{author}{\bibfnamefont{F.}~\bibnamefont{Karsch}},
\bibinfo{author}{\bibfnamefont{P.}~\bibnamefont{Petreczky}},
\bibnamefont{and}
\bibinfo{author}{\bibfnamefont{F.}~\bibnamefont{Zantow}},  
  \bibinfo{journal}{Phys. Lett.\ B} \textbf{\bibinfo{volume}{543}},
  \bibinfo{pages}{41} (\bibinfo{year}{2002}). 

\bibitem[{\citenamefont{Hatsuda and Kunihiro}(1985)}]{Hatsuda2}
\bibinfo{author}{\bibfnamefont{T.}~\bibnamefont{Hatsuda}},
\bibnamefont{and}
\bibinfo{author}{\bibfnamefont{T.}~\bibnamefont{Kunihiro}},  
  \bibinfo{journal}{Prog. Theor. Phys. Phys.} \textbf{\bibinfo{volume}{74}},
  \bibinfo{pages}{765} (\bibinfo{year}{1985}). 

\bibitem[{\citenamefont{Barducci et al.}(2006)}]{Barducci}
\bibinfo{author}{\bibfnamefont{A.}~\bibnamefont{Barducci}}, 
\bibinfo{author}{\bibfnamefont{R.}~\bibnamefont{Casalbuoni}}, 
\bibinfo{author}{\bibfnamefont{G.}~\bibnamefont{Pettini}},
\bibnamefont{and} 
\bibinfo{author}{\bibfnamefont{R.}~\bibnamefont{Gatto}}, 
\bibinfo{journal}{Phys. Lett.\ B} \textbf{\bibinfo{volume}{301}},
\bibinfo{pages}{95} (\bibinfo{year}{1993}). 

\bibitem[{\citenamefont{Karsch et al.}(2001)}]{Karsch}
\bibinfo{author}{\bibfnamefont{F.}~\bibnamefont{Karsch}},
\bibinfo{author}{\bibfnamefont{E.}~\bibnamefont{Laermann}},
\bibnamefont{and}
\bibinfo{author}{\bibfnamefont{A.}~\bibnamefont{Peikert}},  
 \bibinfo{journal}{Nucl. Phys.\ B} \textbf{\bibinfo{volume}{605}},
\bibinfo{pages}{579} (\bibinfo{year}{2001}). 



\end{thebibliography}
\end{document}